# New frontiers in characterising ZrB$_2$-MoSi$_2$ ultra-high temperature ceramics

Mainak Saha

**Abstract**

The structure-property(S-P) correlation, especially in the context of high-temperature oxidation, in ZrB$_2$-MoSi$_2$ based Ultra High-Temperature Ceramic Matrix Composites (UHTCMCs) have been extensively investigated for quite some time since the last 25 years and a countless amount of published data is presently available in this field. On the other hand, emergence of correlative microscopy [1] has completely revolutionised the world of materials research, in a number of ways. However, owing to the challenges of sample preparation, there has hardly been a literature aimed at understanding the aforesaid phenomenon of understanding S-P correlation, based on oxidation in ZrB$_2$-MoSi$_2$ UHTCMCs, which, in future, may open up new frontiers of research in UHTCMCs. The present review intends to discuss some of the most interesting data published existing in this field and intends to provide a brief overview of the challenges associated along with some of the currently unexplored avenues in this field, especially in terms of fundamental research. However, in view of the enormous amount of research already done in understanding oxidation-based S-P correlation in these materials, the author does not claim to address all the issues which may be associated with understanding the same.

**1. Introduction**

Since the early 2000s, the search for materials which may withstand extreme conditions of temperature, chemical reactivity, mechanical stress, radiation, and ablation, especially in the context of hypersonic and space aviation (Re-entry vehicles, most presently) [2-11], has been a driving force for an enormous volume of research based on the borides and carbides of a number of transition metals (such as, Zr, Hf etc.). ZrB$_2$ can be considered a leading material in this field of research owing to a unique combination of properties such as melting point above 3000°C, relatively low density (than the other diborides), high thermal conductivity coupled with good mechanical strength and remarkable refractoriness at elevated temperatures [11-16], ZrB$_2$ has been the centre of attraction in this field. On the contrary, Sinterability of ZrB$_2$ is extremely low, owing to the presence of strong covalent bonds and low self-diffusivity [16], and therefore it requires pressure-assisted sintering techniques at extremely high temperatures (nearly~ 1900˚C) in order to eliminate porosity [17]. Furthermore, the $ZrB_2$ exhibits poor oxidation resistance [16] and mechanical properties [17, 18] owing to the fact that powder surface is always associated with a number of oxides, mainly: $B_2O_3$ and $ZrO_2$, leading to an evaporation-condensation mechanism, which subsequently causes mass transfer without effective densification at low temperature and abnormal grain growth at high temperatures, both resulting in poor mechanical properties [17, 18]. This is due to the reason that formation of glassy $B_2O_3$ and crystalline $ZrO_2$ upon exposure in air at temperatures above 800˚C [16, 17, 19], due to reaction (1):

$$ZrB_2 + 5/2 O_2 = ZrO_2 + B_2O_3 \qquad (1)$$

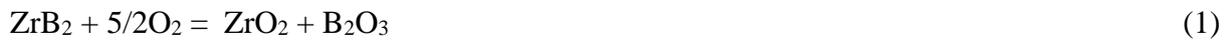

With increase in temperature above 1100˚C, the evaporation of protective glassy $B_2O_3$ layer on $ZrO_2$ occurs, thereby leaving behind a porous $ZrO_2$ scale. With further increase in temperature, the porous $ZrO_2$ scale shows a change in grain morphology: from equiaxed to columnar grains, thus aiding the oxidation process [20, 21]. Effective approaches for simultaneous improvement in sinterability, mechanical properties [16, 18-20] and oxidation resistance of these borides, based on the incorporation of additives have been developed [17-19, 21-25] among which $MoSi_2$ has been reported to be one of the most beneficial, to enhance the densification of $ZrB_2$ at ~1750˚C, owing to the formation of transient liquid phases [26-28]. The formation of glassy $SiO_2$ layer [23, 29] as an effective oxygen diffusion barrier, coupled with the formation of phases such as $MoB$ and $Mo_5Si_3$, and the suppression of volatile species [23], may be attributed to be responsible for the beneficial effect of $MoSi_2$ during oxidation. These aforementioned factors have been reported to be additionally responsible for hindering the formation of columnar $ZrO_2$ grains in the subsurface. This hinders the inward oxygen diffusion into the ceramic matrix [20, 25]. The oxidation reactions of $ZrB_2$ - $MoSi_2$ UHTCMCs have been shown [16, 23, 29-32]:

$$ZrB_2 + 5MoSi_2 + 19/2\ O_2 = ZrO_2 + 7SiO_2 + Mo_5Si_3 + B_2O_3 \qquad (2)$$

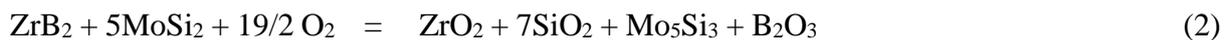

$$2MoSi_2 + B_2O_3 + 5/2 O_2 = 4SiO_2 + 2MoB \qquad (3)$$

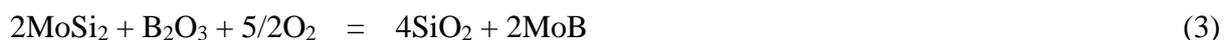

However, unfortunately, the influence of different interfaces, for instance, Grain boundaries (GBs), Interphase Boundaries (IBs) etc. on the grain growth kinetics and during change in grain morphology at elevated temperatures in $ZrB_2$, still needs a lot of experimental understanding, in order to be able to come up with a novel methodology of interface-controlled engineering of microstructures in these UHTCs and UHTCMCs during oxidation at elevated temperatures (> 1900˚C). The novel concept of "Correlative Microscopy", developed a few years back, may be a tool for future investigation of 'structure-property (S-P)' correlation in these materials and coupled with the 1st principle calculations, may be extensively used to address a number of presently unexplored fundamental issues, associated with high-temperature oxidation of these materials. This will be discussed in greater details in the upcoming section.

## 2. Oxidation behaviour of pressureless sintered $ZrB_2$-$MoSi_2$ ceramics based on Sciti et al. [33]

### 2.1 Between 800 and 1000˚C

Based on the report of Sciti et al. [33], the rate of formation of protective silica (from reaction (4)), between 800 and 1000 °C, is not high enough to prevent rapid oxidation. Therefore, the oxidation of $ZrB_2$ to zirconia (reactions (5) and (6)), along with the formation of volatile $MoO_3$ by reaction (4), has been reported to be the main phenomenon in this temperature regime [34, 35].

$$MoSi_2(s) + 7/2 O_2(g) = SiO_2(s) + MoO_3(s) \tag{4}$$

At these temperatures, $ZrO_2$ and liquid $B_2O_3$ are semi-protective and protective, respectively, as confirmed by the kinetic curves, which resemble those for the oxidation of pure $ZrB_2$ [34] in this temperature regime. The consequent volume expansion due to the two aforesaid phenomenon, combined with the escape of gases, leads to formation of voids and cracks in the oxide layer. The presence of crystalline MoB during oxidation at 1000˚C, indicates the occurrence of reaction (5) besides reactions (3) and (4) [35].

$$ZrB_2(s) + 5/2 O_2(g) = ZrO_2(s) + B_2O_3(l) \tag{5}$$

$$B_2O_3(l) = B_2O_3(s) \tag{6}$$

$$ZrB_2(s) + 2MoSi_2(s) + 5O_2(g) = ZrO_2(s) + 2MoB(s) + 4SiO_2(s) \tag{7}$$

Reaction (7) has been reported to be thermodynamically favoured at all temperatures between 700 and 1400˚C, with highly negative values of ΔG at 700 and 1400 °C [35]. However, owing to the slow formation of $SiO_2$, especially at 1000 °C, molybdenum oxide [36–40] and liquid $B_2O_3$ may be expected to expected to undergo some amount of vaporisation.

### 2.2 Between 1000 and 1200˚C

Based on Sciti et al. [33], the production of $SiO_2$ due to oxidation of $MoSi_2$ has been reported to occur at a rate, high enough to prevent inward oxygen transport from the atmosphere. At ~ 1200 °C, formation of cracks in $ZrB_2$, evaporation of liquid $B_2O_3$ has been reported to occur. However, this phenomenon is highly time-dependent viz. after ~ 100 min, a drop in the weight gain, caused due to formation of a continuous silica-rich layer, exhibiting protective action, has been reported [33] wherein, formation of $ZrO_2$ and oxides of Mo, favouring the penetration of oxygen and, consequently, its contact with $ZrB_2$ along $MoSi_2$ particles also in the bulk, have been reported to be responsible for the formation of crack in $ZrB_2$ at ~1200°C. At the interface between the oxide and the unreacted bulk, owing to low partial pressure of oxygen, selective oxidation of Si occurs; resulting in $MoSi_2$ getting depleted of Si and reaction of Mo with B from $ZrB_2$ to form MoB. A very little amount of borosilicate glass, however, has been reported to be observed in the oxide layer, unlike that in $ZrB_2$–SiC composites [35] and besides, formation of MoB through reaction (5) has also been reported to be thermodynamically and kinetically more favourable than the formation of borosilicate glass. According to previous studies [36] when $ZrO_2$ grains are embedded in an $SiO_2$ layer, diffusion and dissolution of interstitial silicon into crystalline zirconia occurs until the solubility limit is attained; after which zircon precipitates out, thereby leading to the formation of zirconia grains, surrounded by a $ZrSiO_2$ shell, further, embedded in the amorphous silica, as shown in reaction (8) [34]:

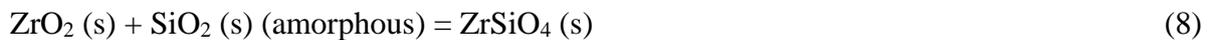
$$ZrO_2 (s) + SiO_2 (s) (amorphous) = ZrSiO_4 (s) \tag{8}$$

The nearly parabolic kinetics of oxidation at 1200 and 1300 °C and the pure parabolic kinetics found at 1400 °C have been reported to be the prime indication for diffusion of oxygen through the silica-rich oxide layer, similar to $ZrB_2$ –SiC UHTCMCs.

**2.3. Other notable reports on oxidation of $ZrB_2$-$MoSi_2$ ceramics**

**2.3.1 Between 700 and 1200°C**

$ZrB_2$ is well-known to oxidize according to the reaction [41, 42]:

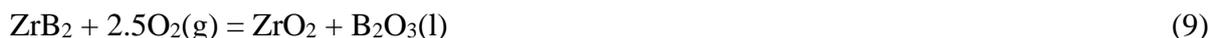
$$ZrB_2 + 2.5O_2(g) = ZrO_2 + B_2O_3(l) \tag{9}$$

The low melting point (450 °C) and high vapor pressure reported for $B_2O_3$ make it undergo vaporisation at low temperatures:

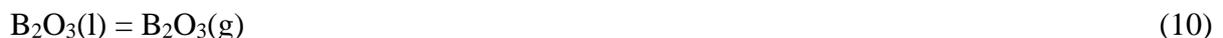
$$B_2O_3(l) = B_2O_3(g) \tag{10}$$

At temperature below 1100 °C, the presence of $B_2O_3$ in the form of a liquid has been reported to hinder oxygen transport [41], thereby following parabolic kinetics [20], while at temperatures higher than 1100°C, the protective action of $B_2O_3$ fades owing to the significant vaporisation of $B_2O_3$, thereby following paralinear kinetics [20]. Formation of $MoO_3$ from $MoSi_2$ has already been shown in Reaction (4). Figs. 1, 2 and 3 show the Scanning Electron Microscopy (SEM) images of $ZrB_2$-20 vol.% $MoSi_2$ oxidised at 700, 1000 and 1200°C, respectively.

$MoO_3$ with high vapor pressure is volatile between 500 and 800 °C and melts at ~ 801 °C [42]. At temperatures higher than 1200 °C, $SiO_2$ tends to act as a barrier against oxidation, whereas, at lower temperatures, formation of $MoO_3$, at a much higher rate than that of $SiO_2$, [43, 44] tends to dominate the oxidation of monolithic $MoSi_2$, as a result of which there is a volume expansion leading to the build-up of a huge amount of internal stresses at grain boundaries (GBs).

The production of silica between 700 and 1000 °C has been reported to be slow enough for protecting the surface [30, 41]. With an increase in oxidation temperature above 1000 °C, the formation of MoB and ZrC secondary phases, mainly concentrated near the sample surface, has been reported during sintering [41]. Oxidation of MoB has been reported to occur, according to Reaction (11) [30, 41, 45]:

$$2MoB + 3.5O_2(g) = B_2O_3 + 2MoO_2 \text{ (l)} \tag{11}$$

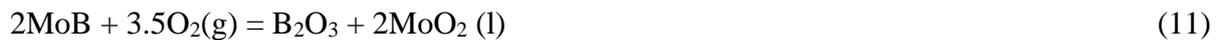

The oxide growth leads to ZrC undergoing catastrophic oxidation at relatively low temperatures, resulting in high stress levels along GBs [46]. Oxidation of ZrC has been reported to occur based on Reaction (12)

$$ZrC + 2O_2(g) = ZrO_2 + CO_2(g) \tag{12}$$

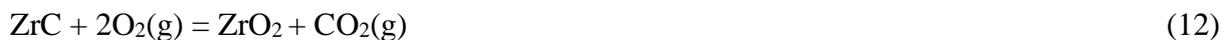

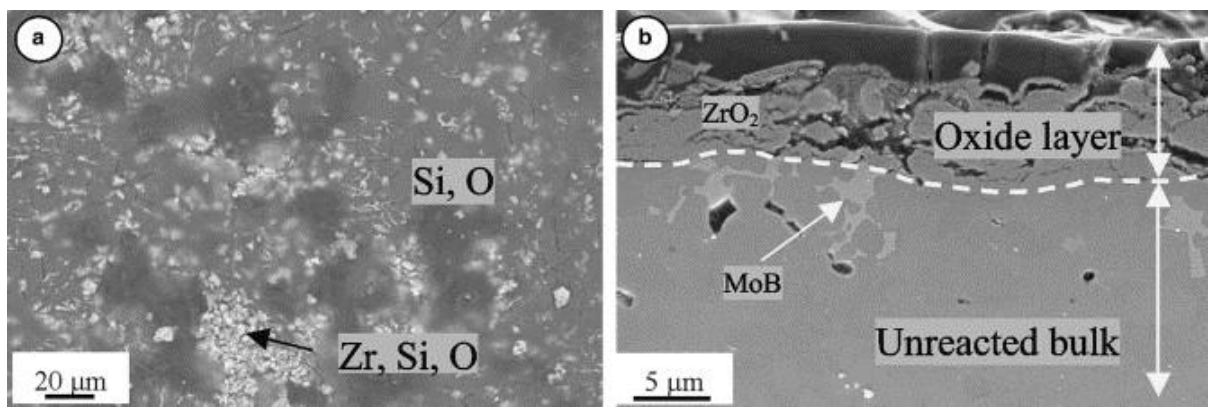

**Fig. 1** SEM images showing **(a, b)** Surface and **(c)** Cross-section of $ZrB_2$ –20 vol.% $MoSi_2$ oxidised at 700°C [41].

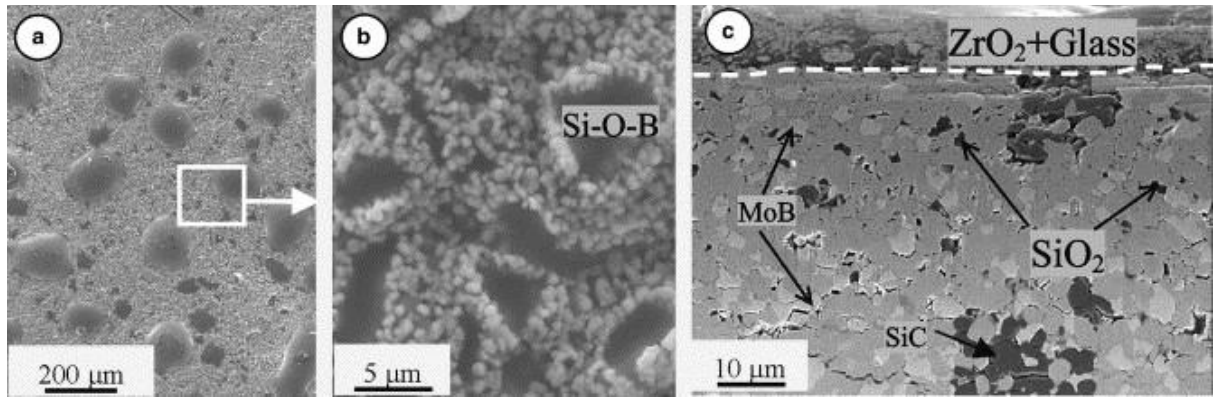

**Fig. 2** SEM images showing **(a, b)** Surface and **(c)** Cross-section of ZrB2 –20 vol.% $MoSi_2$ oxidised at 1000°C [41].

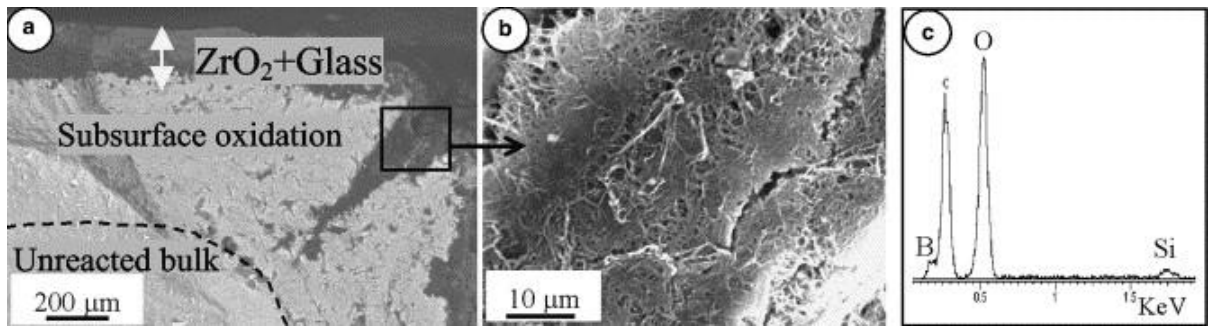

**Fig. 3** SEM images showing **(a, b)** Surface and **(c)** Cross-section of ZrB2 –20 vol.% $MoSi_2$ oxidised at 1200°C [41].

### 2.3.2 Between 1200 and 1400˚C

Based on Silvestroni et al. [30] and Sciti et al. [41], the formation of $SiO_2$, as shown in reaction (4), may be expected to be rapid enough to protect the $ZrB_2$-$MoSi_2$ UHTCMC. It was hypothesized in the reports [30, 41] that at the early stage of oxidation of the UHTCMC, $SiO_2$ may not have been completely formed and that the Reactions (4) and (9)–(12) led to an extremely rapid rate of degradation and this process is also aided by the presence of fast-oxidizing species such as MoB and ZrC. The formation of cracks, in a way, is detrimental, not only from the mechanical point of view but also because cracking leads to rapid transport of oxygen to the internal regions in the sample and this subsequently, leads to an increase in the area exposed to oxidation. Liquid $B_2O_3$ has been reported to react with $SiO_2$, thereby forming a borosilicate glass, similar to that observed for $ZrB_2$–SiC composites [47-49] and which, owing to low viscosity, may get easily penetrated into the bulk of $ZrB_2$-$MoSi_2$ sample, through voids left behind by the oxidation of $ZrB_2$.

Based on the work of Sciti et al. [41], formation of cracks and the presence of fast oxidizing species such as ZrC and MoB in the sub-surface of the sintered samples leading to rapid sub-surface oxidation have been reported to be the main factors for controlling the degradation of $ZrB_2$-$MoSi_2$ UHTCMC between 1200 and 1400 ˚C.

On the other hand, based on the work of Silvestroni et al. [30], consumption of of $B_2O_3$ content in the subsurface liquid through reaction with $MoSi_2$ to form solid MoB [50], may be attributed to the retardation in formation of $ZrO_2$, on adding $MoSi_2$. In addition, the dissolution of a small amount of Mo into the $SiO_2$ scale, which has actually been reported to act as a reticulating agent in glass structures [51, 52], may increase the viscosity of the liquid phase, thereby acting as an effective barrier against inward diffusion of oxygen from the atmosphere to the sample surface. Besides, the pronounced bubbling activity on the surface of the oxidized sample at 1800 ˚C, reported by Silvestroni et al. [30], indicates that various gaseous species are formed in the subsurface layers. Interestingly, during the oxidation of $ZrB_2$-$MoSi_2$ UHTCMC, MoB has been reported to occur in 2 morphologies, [27, 30, 41] depending on the oxidation temperature. At ~1650 ˚C MoB is homogeneously spread and presently mostly in nano-inclusions within zirconia grains, whereas, at 1800 ˚C, a thick cordillera ~5mm is observed.

From the mechanical point of view, degradation caused due to oxidation of ZrB2 has been reported to be responsible for the massive drop in flexural strength in the material at 1300 and 1400˚C [41]. Till date, there is limited data available for $ZrB_2$-$MoSi_2$ UHTCMCs, for oxidation at temperatures exceeding 1400 ˚C, although considerable amount of work has been done in investigating phase evolution during oxidation at 1650 and 1800 ˚C by Silvestroni et al. [30].

The major conclusion from the work of Silvestroni et al. [30, 53] is that a homogeneous distribution of Mo-rich phases, i.e., a continuous (Zr, Mo)$B_2$ solid solution without large Mo-phases, is desirable for enhancing the oxidation resistance of $ZrB_2$-$MoSi_2$ UHTCMCs.

## 3. Criterion for design of UHTCMCs with improved oxidation resistance

The choice of cation to be dissolved into the boride matrix is of extreme importance. As long as the nano-inclusions within $ZrO_2$ grains are located at oxygen partial pressure values of below $10^4$-$10^6$ MPa between 1650-1850°C (Fig. 4), they may be considered to be stable compounds [30]. Hence, the main criterion, to be focussed on for the design of UHTCMCs should be based on using suitable additives (Me) with $ZrB_2$, in order to form a homogeneous (Zr, Me)$B_2$ solid solution with an extremely high phase stability in the region of the highest partial pressure of oxygen, without any amount of GB segregation, whatsoever. Based on the Predominance diagrams at different temperatures, Mo, Cr and W have been reported to possess the highest stability field at the highest oxygen partial pressures, [30, 53] in-line with the experimental findings of Kazmedashi et al. [54] and Talmy et al. [55].

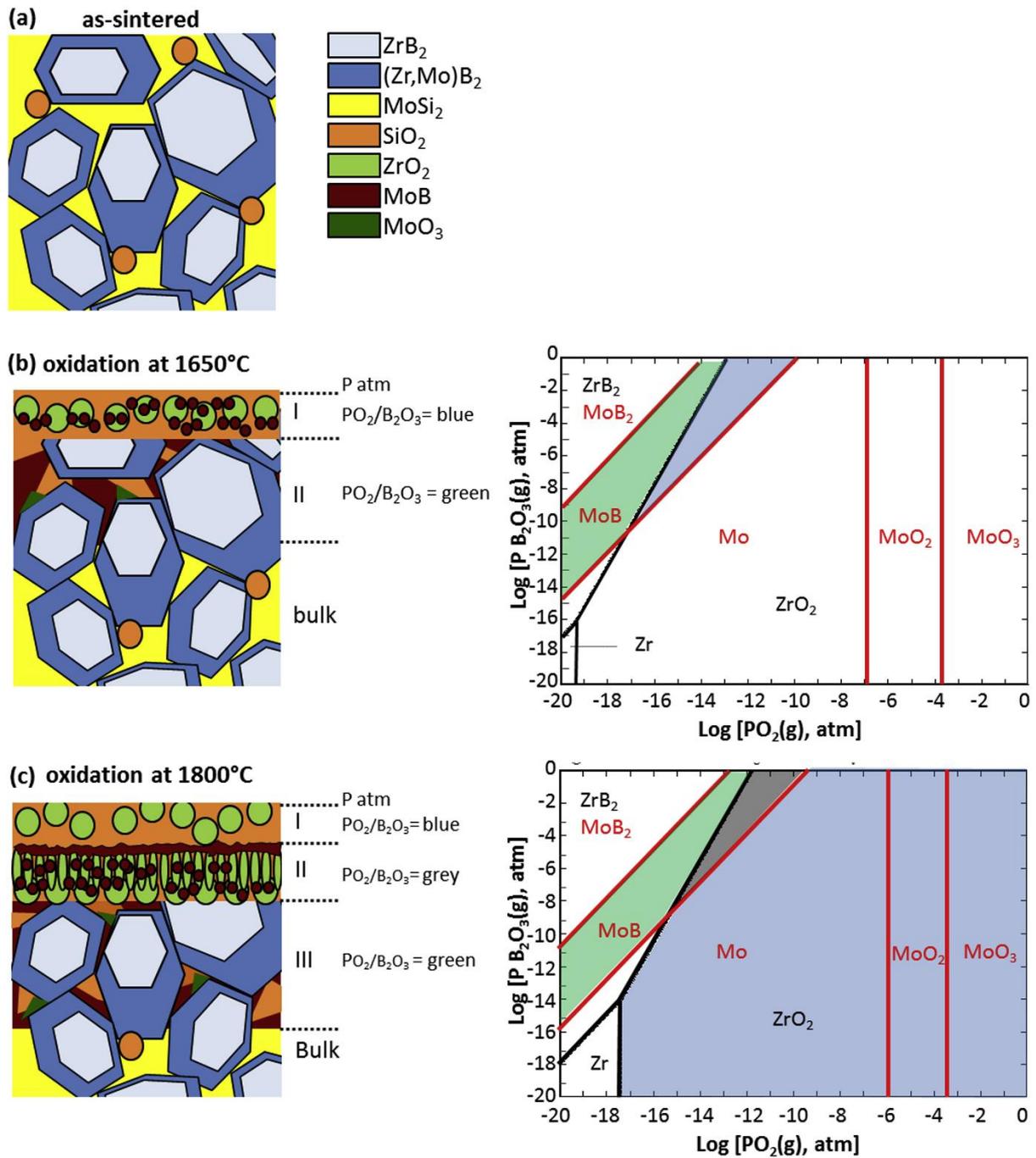

**Fig. 4** Left-hand side: Schematic of microstructural evolution starting from **(a)** as-sintered state up to oxidation at **(b)** 1650 and **(c)** 1800 °C. Right-hand side: phase stability diagrams for the Zr-B-O (black) and Mo-B-O (red) system as a function of partial pressures of oxygen and boron oxide at **(b)** 1650 °C and **(c)** 1850 °C [30].

## 4. Anisotropy in oxidation of hot-forged (textured) ZrB$_2$–MoSi$_2$ UHTCMCs

Liu et al. [56] investigated the anisotropic oxidation behaviour of textured ZrB$_2$–MoSi$_2$ UHTCMCs, prepared by reactive hot pressing and subsequent hot forging, for different exposure times(ranging from 0.5h to 12h) at 1500°C and reported that on exposure in air for a relatively short period of time (0.5 h and four h), the ceramics, textured as a result of the aforesaid deformation process, exhibit a good amount of anisotropy in oxidation i.e., the V-direction surface has a much better oxidation resistance than the P-direction surface (Fig. 5). The V-direction surface of the texture ceramics also shows a better oxidation resistance, in contrast to the V-direction surface in non-textured ceramics, which may be certainly attributed to the higher atomic density of $ZrB_2$ grains and a much higher diffusion path length in the V- than in the P- direction. However, with passage of time, the formation of thick $SiO_2$ layer, preventing the inward diffusion of oxygen from the outer atmosphere, has been reported to weaken the anisotropy in oxidation. Liu et al. [56] have finally concluded that the combination of good oxidation resistance and flexural strength in the V-direction surface for textured $ZrB_2$–$MoSi_2$ UHTCMCs may enable these materials to act as promising candidates for future applications, especially for improving the performance of nosecones of the most trending hypersonic re-entry vehicles.

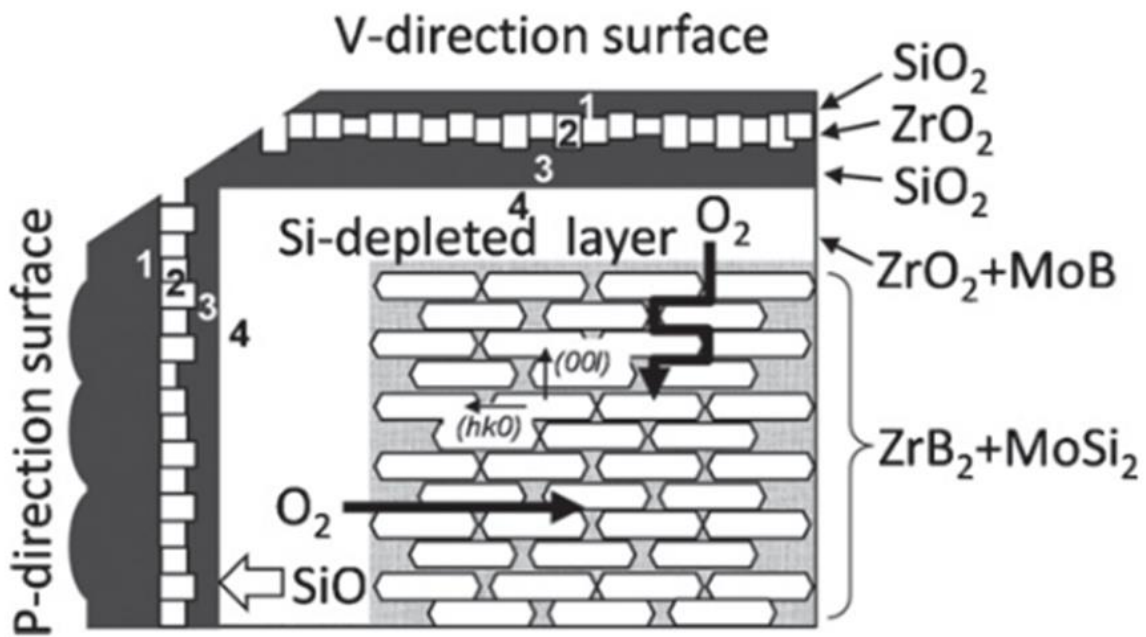

**Fig. 5.** Schematic representation of both V- and P-direction cross sections in textured $ZrB_2$–$MoSi_2$ ceramics oxidized for four h at 1500°C, illustrating anisotropy in oxidation behaviour [56].

## 5. Mechanical properties of ZrB₂-MoSi₂ ceramics due to oxidation

Grohsmeyer et al. [57] have reported that the Elastic properties (Fig. 6(a)) and Vickers microhardness (Fig. 6(b)) of $ZrB_2$-$MoSi_2$ ceramics follows rule of mixture trends based on the constituent phases and that the Young's modulus also decreases with decreasing $ZrB_2$ average particle size due to the presence of higher amounts of low modulus tertiary oxide phases such as $SiO_2$ in compositions prepared from finer $ZrB_2$ powders and is highly coherent with the previous experimental investigations [58-68]. Besides, the Flexural strength at 1500 °C for $MoSi_2$ contents above 25 vol.% has been reported to vary slightly between 420 and 460 MPa for all compositions, whilst for lower contents, two trends were reported [57]. For fine and medium $ZrB_2$ grades, flexural strength was reported to increase with MoSi2 content due to a probable transition from unprotective to protective mechanism of oxidation. Besides, for these compositions, a continuous borosilicate glassy scale was observed and this attributed as the main factor for limiting the depth of oxidation damage. For coarse $ZrB_2$ particles, flexural strength was reported to decrease with increasing $MoSi_2$ content owing to the ductilization of the disilicide, thereby leading to the partial masking of the beneficial effect of absorbing fracture energy possessed by the solid solution shell [57].

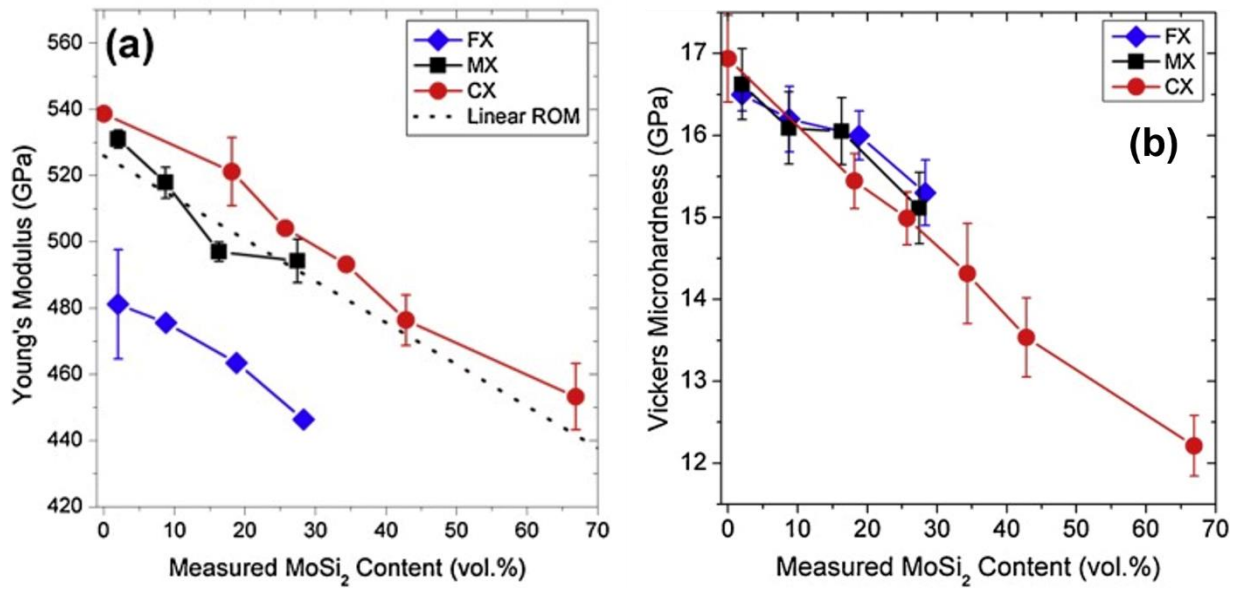

**Fig. 6 (a)** Young's modulus and **(b)** Vickers microhardness as a function of $MoSi_2$ content in $ZrB_2$-$MoSi_2$ ceramics. N, M and C indicate coarse, medium and fine powder grades of $ZrB_2$, respectively. X indicates vol.% of $MoSi_2$. In part **(a)**, ROM refers to volumetric rule of mixtures [57].

## 6. A few presently unexplored avenues in this field and outlooks for the future

The onset of correlative microscopy [1] has completely revolutionised the field of research in structural materials. Ranging from the study of segregated atoms [4, 5, 10] to characterising nanoscale precipitates at different interfaces [5], the novel technique has enabled material scientists to correlate the properties of a material with its bulk and interfacial structure, simultaneously. From the fundamental viewpoint, there still remain a lot of unexplored avenues, in the context of oxidation studies in the UHTCMCs, which may be potentially addressed by this novel technique. For instance, in the context of oxidation of $MoSi_2$ in $ZrB_2$-$MoSi_2$ based UHTCMCs at temperatures lower than 1200 °C, formation of $MoO_3$, has been reported by Silvestroni et al. [30], to occur at a rate which is much lower than that of $SiO_2$, thereby leading to a volume expansion and a consequent piling up of a huge amount of internal stresses at grain boundaries (GBs). Unfortunately, at present, there is limited understanding of the effect of volume expansion associated with the formation of $MoO_3$ at temperatures lower than 1200 °C on the stress pile-up at GBs. Hence, its the GB cohesion [10] and GB energy [5] which needs to be addressed, for a detailed understanding of the influence of the same on the stress pile-up associated with volume expansion, due to formation of $MoO_3$, in the present context. Besides, adhesion between the oxide layer and the surface of the base material ($ZrB_2$-$MoSi_2$, in the present context) is also something which needs to be addressed as a function of the local chemistry at the substrate-oxide interface before subjecting these materials to service at elevated temperatures.

Although a number of theoretical studies based on the kinetics of oxidation as a function of the diffusivity of oxygen, have illustrated the properties of the oxides, formed during oxidation of $ZrB_2$-$MoSi_2$ at different temperatures [26-30], but however, there exists a huge possibility of understanding the same experimentally, through detailed study of oxygen-defect interaction, using extensive defect and orientation based microscopy techniques such as Electron Backscatter Diffraction (EBSD) [10], Electron Channeling Contrast Imaging (ECCI) [69] e.t.c. and correlating the structural information, obtained henceforth with the atomic-level chemistry using Atom Probe Tomography (APT) [7] technique, using Focussed Ion Beaming (FIB) [7] as a lift-out technique to prepare samples for APT. This, in a way, may also be used a means to understand the change in GB character [5] of different crystalline oxides formed during oxidation at different temperatures and determining there is a localised phase formation at GBs, due to the same, as recently reported by Meiners et al. [70]. Addressing a couple of fundamental issues, as these may, henceforth be used as a pre-requisite to design $ZrB_2$-$MoSi_2$ based

UHTCMCs, in future, with even better mechanical properties during oxidation at elevated temperatures.

## 7. Conclusions

There is no doubt that $ZrB_2$-$MoSi_2$ UHTCMCs will find numerous applications in future, owing to the excellent combination of physical, mechanical, and electrical properties of these materials, and that the research associated with these materials, would be aimed at further improvement in different properties of these materials, however, exploring the currently unaddressed avenues, would be the pre-requisite for aiming at property enhancement based on understanding structure-property correlation at different oxidation conditions in these materials.

**Acknowledgement**

Mr. Mainak Saha is grateful to the Department of Metallurgical and Materials Engineering, NIT Durgapur for extending its valuable support while carrying out the present review.

**Conflict of Interest**

The author hereby declares no conflict of interest.